\documentclass[journal]{IEEEtran}
\ifCLASSINFOpdf
\else
\fi
\def\ps@headings{%

\def\@oddhead{\mbox{}\scriptsize\rightmark \hfil \thepage}%

\def\@evenhead{\scriptsize\thepage \hfil \leftmark\mbox{}}%

\def\@oddfoot{}%

\def\@evenfoot{}}

\makeatother

\usepackage{multirow}
\usepackage{amsfonts}
\usepackage{amsmath}
  \usepackage{graphicx}
  \usepackage{algorithm}
\usepackage{setspace}
\usepackage{subfigure}
\usepackage{array}
\newcommand{\PreserveBackslash}[1]{\let\temp=\\#1\let\\=\temp}
\newcolumntype{C}[1]{>{\PreserveBackslash\centering}p{#1}}
\newcolumntype{R}[1]{>{\PreserveBackslash\raggedleft}p{#1}}
\newcolumntype{L}[1]{>{\PreserveBackslash\raggedright}p{#1}}
\begin{document}
%
\title{Joint Source-Channel Coding for Real-Time Video Transmission to Multi-homed Mobile Terminals}

\author{Xiaoyan~Gao\\
School of Computer Science and Technology, North China Institute of Science and Technology\\
Email:  xiaoyan\underline{ }gao@ncist.edu.cn.

}



%


\maketitle

\begin{abstract}
This study focuses on the mobile video delivery from a video server to a multi-homed client with a network of heterogeneous wireless. Joint Source-Channel Coding is effectively used to transmit video over bandwidth-limited, noisy wireless networks. But most existing JSCC methods only consider single path video transmission of the server and the client network. The problem will become more complicated when consider multi-path video transmission, because involving low-bandwidth, high-drop-rate or high-latency wireless network will only reduce the video quality. To solve this critical problem, we propose a novel Path Adaption JSCC (PA-JSCC) method that contain below characters: (1) path adaption, and (2) dynamic rate allocation. We use Exata to evaluate the performance of PA-JSCC and Experiment show that PA-JSCC has a good results in terms of PSNR (Peak Signal-to-Noise Ratio).

\end{abstract}
\begin{keywords}
mobile video delivery; heterogeneous wireless networks; multi-homing; joint source channel coding
\end{keywords}


%
\IEEEpeerreviewmaketitle

\section{Introduction}
Nowadays handheld devices (e.g., smartphone and iPad) is explosive increasing, on the other hand,  mobile video streaming (e.g., Hollywood [1] and Metacafe [2]) has became one of the most popular portion of network flow. According to the Cisco Visual Index [3] report, the percentage of mobile video streaming is $57\%$ in 2012 and will increase to $69\%$ by the year 2017. As a result, focus on mobile video quality guarantee study is meaningful.

With the progress of technology mobile devices users could access network with various options (e.g., WLAN, 3G/4G and WiMAX), but problem still existing because bandwidth limitation of single wireless causing bottleneck of mobile video quality promotion. Contemporary WLAN systems could not guarantee video streaming quality because of small coverage area and limited bandwidth when there are number of mobile users [4-5]. Meanwhile, cellular networks (e.g., UMTS) could provide more reliable service to mobile users, but their bandwidth is finite for high-quality video transmission [6]. The 4G LTE and WiMAX could provide higher transmission rate and larger coverage area, however, they are not widely applied and bandwidth limitation also existing as the number of mobile users increases. Above all, the limitation of single wireless network turn research attentions to multi-path video transmission.

To solve the existing problem, researchers proposed joint source-channel coding (JSCC) and proved it was an effective approach in noisy wireless networks. But existing JSCC approaches (e.g., [7-8]) only consider single path between server and client to transfer mobile video. The problem becomes more complicated when consider multi-path video transmission in heterogenous wireless network. Fig. 1 shows the problem, in location 1, the user encounter a video problem because the cellular link not transfer the streaming well. In location 2, the user requests to the video server and in the meanwhile access WLAN network, but because of the unstable WLAN link, the video quality reduce more. In location 3 which the place user could connect to WiMAX and then the perceived quality became better than in location 1 and 2. This can be summarized as the user-perceived video quality will only be reduced when involving an unreliable wireless access network.

\begin{figure}
\centering
 \includegraphics[width=0.5\textwidth,keepaspectratio]{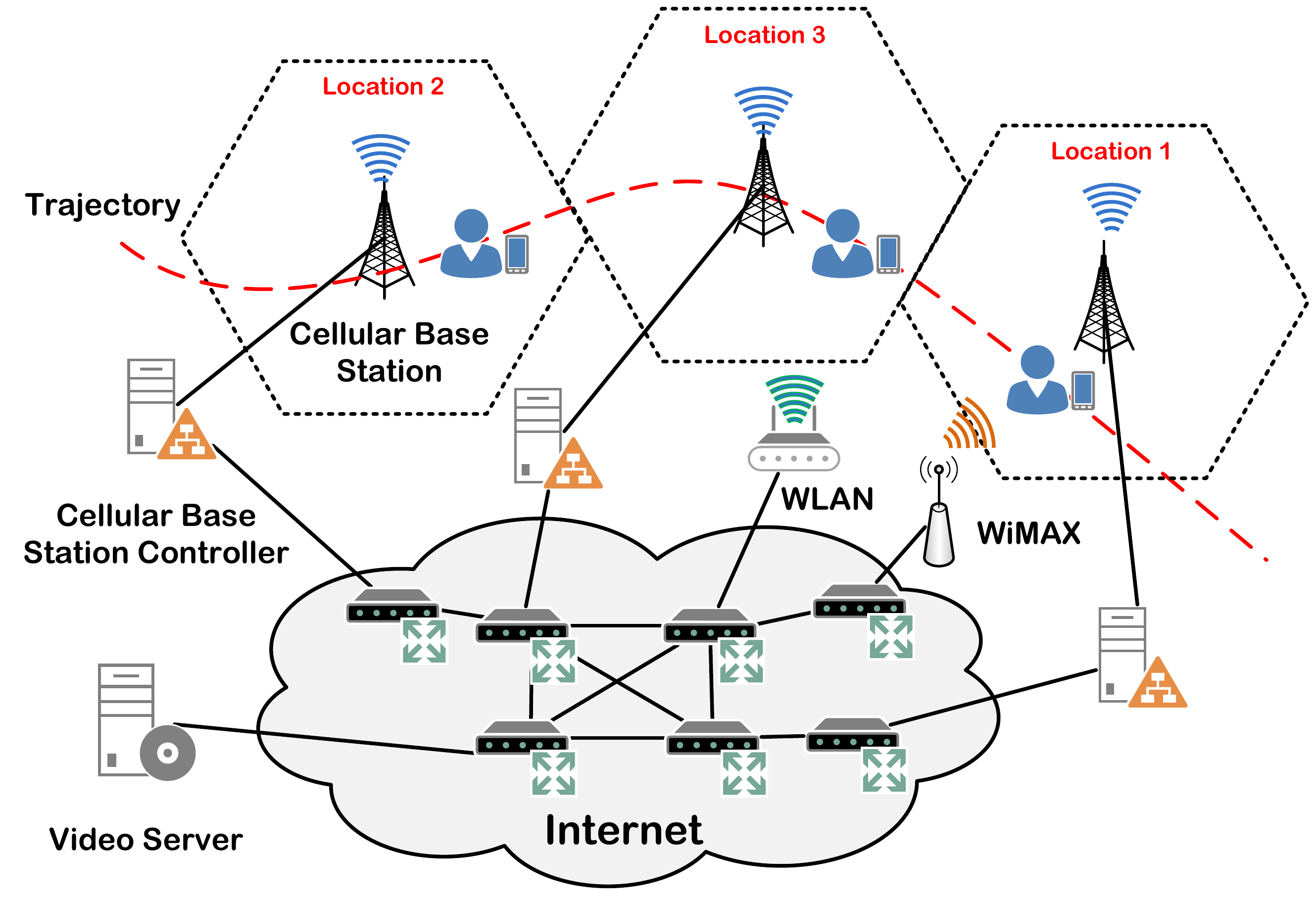}
 \caption{Illustration of a mobile video streaming service in a heterogeneous wireless network.}\label{fig1}
 \vspace{-5pt}
\end{figure}
To improve the JSCC approach for mobile video transmission in heterogeneous wireless networks, we propose a path adaption JSCC (PA-JSCC) in this study. The term 'path adaption' means dynamically selecting the appropriate wireless path and allocating the appropriate data rate to each of them. First, the video source rate adaption scheme is designed to meet the delay requirements of real-time video transmission. Second, the Forward Error Correction (FEC) redundancy estimation is employed to decrease the tolerable loss rate. Third, a effective flow allocation algorithm is proposed to reduce the end-to-end video distortion.

Specifically, the contributions of this paper can be summarized in the following:
\begin{itemize}
  \item We present an efficient end-to-end video transmission solution which using JSCC and path selection algorithm in heterogeneous wireless networks in order to improve the transmission video quality.
  \item We develop a mathematical model of JSCC for reducing the end-to-end video distortion with multi-path transmission.
  \item Using Exata implements semi-physical emulations with the real-time H.264 video streaming.
\end{itemize}


The remainder of this paper is structured as follows. In Section II, we briefly discuss the related work. Section III presents the system model and problem formulation. The design of the proposed PA-JSCC description in detail is in section IV. In Section V, we provided the performance evaluation. Conclusion remarks are given out in Section VI. The basic notations used throughout this paper are listed in Table I.
\begin{table}[htbp]
\footnotesize
\renewcommand{\arraystretch}{1}
      \caption{Basic notations used in this paper}
      \vspace{-5pt}
        \centering
      \begin{tabular}{|c|l|}
      \hline
       \textbf{Symbol} & \textbf{Definition} \\
        \hline
        \hline
        $\mathbb{P}$ & the probability value.\\
        \hline
        $\mathcal{P},P_{r}$ &  the set of wireless access networks, $r$th access network.\\
        \hline
        $\mathcal{R}$ & the number of available wireless networks.\\
        \hline
        $\mu_{r}$ & available bandwidth of $P_{r}$.\\
        \hline
        $\pi_r$ & the average loss rate of $P_{r}$.\\
        \hline
        $t_r$ & the propagation delay of $P_{r}$.\\
        \hline
        $\pi_B^*$ & the effective loss rate.\\
        \hline
        $1/\xi_B^r$ & the average burst length of $P_r$.\\
        \hline
        $k$ & the number of source packets in a FEC block.\\
        \hline
        $n$ & the total number of FEC packets in a FEC block.\\
        \hline
        $k/n$ & the FEC code rate.\\
        \hline
        $V$ & the video source (encoding) rate.\\
        \hline
        $S$  & the FEC data packet size.\\
        \hline
        $J$ & the number of video frames in a GoP.\\
        \hline
        $F$  & the video frame rate.\\
        \hline
      \end{tabular}
\end{table}
\normalsize

\section{Related Work}
The related work to this study can be classified into two categories: (1) joint source channel coding, and (2) video transmission in heterogeneous wireless networks. In this section, we will discuss on each topic respectively.
\subsection{Joint Source-Channel Coding}
Most study of the joint source channel coding in video transmission are concentrated on flowing items: 1) find out most fitting video data rate for video flow, e.g., [9]; 2) deigning the channel coding algorithm to achieve the target rate, the Reed-Solomon [10], Turbo [11], and Fountain [12] codes; 3) considering the channel condition and design the video coding scheme to satisfied the requirement, e.g., [13]. But all this works only consider single path video transmission, it is a waste of multi-homed mobile devices. Different from previous JSCC methods, [14] study on the optimal FEC scheme and layer selection in multi-path scenario.

\subsection{Video Transmission in Heterogeneous Wireless Networks}
Video transmission in heterogeneous wireless networks has been a hot area of research and summarized in [15-16]. The EMS [17] and MPLOT [18] are typical protocols for heterogenous network based on erasure code. The Encoded Multipath Streaming (EMS) scheme splits traffic loads over multiple paths according to the path loss rate and dynamically adjusts the FEC redundancy. However, EMS was generally under the assumption that all the available paths could be beneficial for the transmission as in [14]. Multi-Path LOss Tolerant (MPLOT) is a transport protocol that aims at maximizing the throughput of the upper-layer application. But MPLOT can not guarantee real-time video delivery as it does not address tight delay constraints.

\section{System Model and Problem Formulation}
The PA-JSCC system model is described in Fig. 2. The scenario we considered is a heterogenous wireless network with $\mathcal{R}$ wireless connections from a video server to a mobile device. This system involves three models: network model, end-to-end video distortion model and forward error correction (FEC) model. Next, we will introduce each of them simply.

\subsection{Network Model}
We consider each physical path $P_r$ is associated with the following metrics to describe the network model:

\begin{itemize}
  \item the available bandwidth $\mu_r$, expressed in the unit of Kbps.
  \item the propagation delay $t_r$, which includes the link delays of the wired and wireless .
  \item the average loss probability $\pi_B^r\in[0,1]$.
\end{itemize}
We employ continuous time Gilbert model to describe the burst loss behavior on physical path. The state $\mathcal{X}_{r}(t)$ assumes one of two values: $G$ (Good) or $B$ (Bad). $G$ means the packet is successfully delivered at time $t$. Otherwise, $B$ means the packet is lost.
We denote by $\pi^{r}_G$ and $\pi^{r}_B$ the stationary probabilities that $P_{i}$ is good or bad. Let $\xi^{r}_B$ and $\xi^{r}_G$ represent the transition probability from $G$ to $B$ and $B$ to $G$, respectively. In this paper, we adopt two system-dependent parameters to specify the continuous time Markov chain packet loss model: (1) the average loss rate $\pi_B^r$, and (2) the average loss burst length $1/\xi_B^r$. Then, we can have:
\begin{equation}
\begin{split}
\pi^{r}_{G} = \frac{\xi_B^r}{\xi_B^r+\xi_G^r},\text{ and }
\pi^{r}_{B} = \frac{\xi_G^r}{\xi_B^r+\xi_G^r}\cdot
\end{split}
\end{equation}

\subsection{Video Distortion Model}
Here we will introduce a generic video distortion model [19]. The end-to-end distortion ($D_{\text{total}}$) include the channel distortion ($D_{\text{chl}}$) and the source distortion ($D_{\text{src}}$). Therefore, the end-to-end distortion can be written as:
\begin{equation}
\begin{split}
D_{\text{total}}=D_{\text{chl}}+ D_{\text{src}}.
\end{split}
\end{equation}
$D_{\text{src}}$ is mostly dependent on the video source rate and the video sequence parameters (e.g., the sequence complexity in proportion to the source distortion for the same encoding bit rate). The channel distortion is determined by the effective loss rate $\pi^{*}_B$, which is caused by the delivery lost and video packets overdue. We can formulate $D_{\text{total}}$ (in units of mean square error, MSE) as:
\begin{equation}
D_{\text{total}}=\underbrace{D_{0}+\frac{\alpha}{V-V_0}}_{D_{\text{src}}}+\underbrace{\beta\cdot\pi_B^{*}}_{D_{\text{chl}}}\,,
\end{equation}
in which $\alpha$, $V_0$, $D_0$ and $\beta$ are constants for a specific video codec and video sequence. Since this model takes into account the effects of intra coding and spatial loop filtering, it provides accurate estimates for end-to-end distortion [20].
\subsection{Forward Error Correction}
In this paper, we employ systematic Reed-Solomon (RS) block erasure code to eliminate channel losses. Generically, a FEC block of $n$ data packets includes $k$ source packets and $n-k$ redundant packets. Most of time the receiver can fully reconstruct the original $k$ data packets if at least $k$ packets of the FEC block are correctly received. 
Particularly, the FEC code rate $n/k$ needs to be dynamically chosen based on the loss requirement and channel status.

In this paper, Reed-Solomon code is adopted to construct a virtual connection by aggregating multiple heterogeneous wireless networks. Actually, the uncertainty of the resulting virtual connection may increase although its total bandwidth is higher as the number of integrated wireless networks increases. Fountain code is very useful to overcome the uncertainty of a virtual path since its code rate is not fixed in a priori way and can be adjusted flexibly according to the wireless channel states. In general, a large source block size can reduce sensitivity to the wireless channel uncertainty at the cost of increased delay. In addition, fountain code has some advantages for video transmission over wireless networks due to its low encoding/decoding processing demand and high coding efficiency.

Actually, the frame-level [21], GoP-level [22] or sub-GoP level [23] FEC coding is often applied for video data protection. In this work, the GoP-level FEC (see Fig. 3) is employed for the data protection.

\subsection{Problem Formulation}
\vspace{-2pt}
We are now ready to formulate the problem of flow rate allocation combining the JSCC for video delivery in heterogeneous wireless networks. Note that, it is not practical for the video encoder to trace the frequent variation in source rate. Therefore, we adapt the source rate in units of GoP, based on the channel status, FEC code rate and delay requirements. To allow fast adaptation of the source rate to abrupt changes in the video content, this parameter is updated for each group of pictures (GOP) in the encoded video sequence, typically once every $0.25$ second (with $J=8$ frames, $F=30$ fps). The objective is to minimize the summation of the total distortion $D_{\text{total}}$ subject to loss, delay and bandwidth constraints.
\begin{displaymath}
\begin{split}
&\text{For each GoP, determine the value of }\Phi,\,\Omega,\,V,\,n\\
&\text{to minimize }\\
&D_{\text{total}}=D_{\text{total}}=\underbrace{D_{0}+\frac{\alpha}{V-V_0}}_{D_{\text{src}}}+\underbrace{\beta\cdot\pi_B^{*}}_{D_{\text{chl}}}\,,
\end{split}
\end{displaymath}
\begin{equation}
\text{subject to:}
\begin{cases}
\begin{split}
&V\cdot n/k\cdot \frac{\omega_r}{\sum_{i=1}^\mathcal{R}\omega_i}<\mu_r,\,\text{for }1\leq r <\mathcal{R},\\
&V\cdot n/k\leq\sum_{r=1}^{\mathcal{R}}\mu_{r},\\
\end{split}
\end{cases}
\end{equation}
This is a nonlinear optimization problem with linear constraints. With regard to the computational cost and convergence, it is impractical to derive the exact solution for the minimal video distortion. In the next section, we will show how to resolve this optimization problem in the design of the proposed PA-JSCC.
\section{Design of Flow Rate Allocation based Joint Source Channel Coding}
In this section, we describe the overall design of the proposed PA-JSCC and outline the functionality of its major components. The architecture of the system framework is presented in Fig. 4. The system design includes components implemented in both the server and client side, respectively. In order to solve the optimization problem, at the server side we add two components: (1) path selector, (2) flow rate allocation. Particularly, the server side include a data distributor components, it responsibilities is split encoded video streaming into multiple available wireless networks and dispatching the FEC data packets onto different channels.

At the client side, the video frames will be stored in the playback buffer after the FEC decoding process. The inter-frame resequence step aims at reordering the video frames in case they arrive at the client out-of-order. As each video frame is associated with a decoding deadline, the overdue frames will be discarded and concealed by copying from the last received ones. And we add an information feedback components to send channel status information back to server side. Next, we will describe the key components in the system design and their major functions.

\begin{figure}[htbp]
\vspace{-5pt}
\centering
 \includegraphics[width=0.5\textwidth,keepaspectratio]{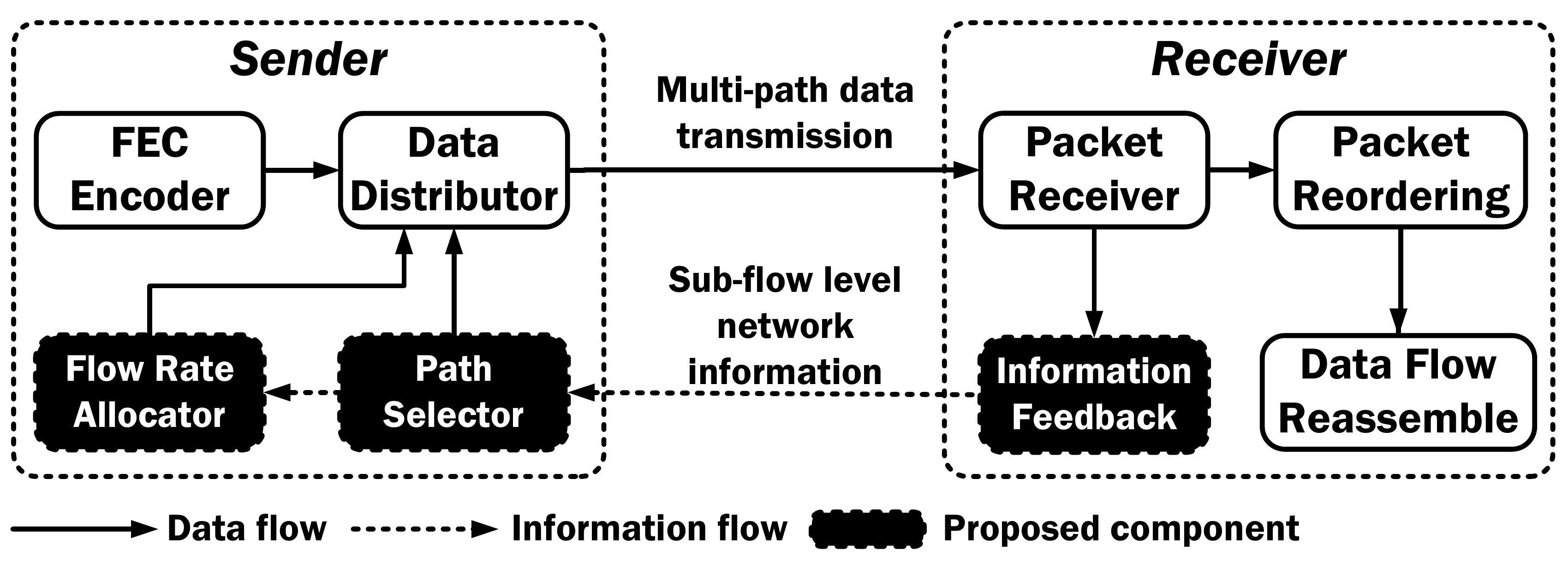}
 \caption{System architecture for performance evaluation.}\label{7}
 \vspace{-10pt}
\end{figure}

\subsection{Path Selector}
According to information theory [24], the video source distortion can be minimized by increasing the effective encoding rate. On the other hand, the increasing encoding rate will lead to higher transmission rate which impose heavier load on channels. If the imposed load exceeds the network capacity, it will in turn result in longer delay and packet loss due to the network congestion. There is a inherent conflict between the source and channel distortion. The functions of path selector component contains determine the code rate on each path and define the maximum number of packets which can be transmitted using the constructed virtual link.
\subsection{Flow Rate Allocator}
The source packets together with the redundancy packets consist of the `flow' mentioned throughout this paper. The goal of the flow rate allocation is to select appropriate wireless access networks out of all the candidates so as to minimize end-to-end video distortion. The functions of the flow rate allocator is to allocate optimal flow rate of the flow.
\subsection{Information Feedback}
Estimating channel status information based on information feedback has been attracting research attentions for years. Over heterogeneous wireless networks, it is very important to identify the physical characteristics of each wireless channel in order to utilize channel resources efficiently. The available bandwidth, propagation delay and channel loss rate are especially important properties for a high quality video streaming service. Information feedback components is used to transfer this necessary information back to server.

\section{Performance Evaluation}

In this section, we evaluate the efficacy of the proposed PA-JSCC by comparing it with the existing schemes for video delivery over heterogeneous wireless networks. We first describe the emulation methodology that includes the emulation setup , performance metric and emulation scenario.
\subsection{Emulation Methodology}
\subsubsection{Emulation Setup}
We adopt the Exata and JSVM as the network emulator and video codec, respectively. The architecture of evaluation system is presented in Fig. 5 and the main configurations are set as follows.
\begin{figure}[htbp]
\vspace{-5pt}
\centering
 \includegraphics[width=0.45\textwidth,keepaspectratio]{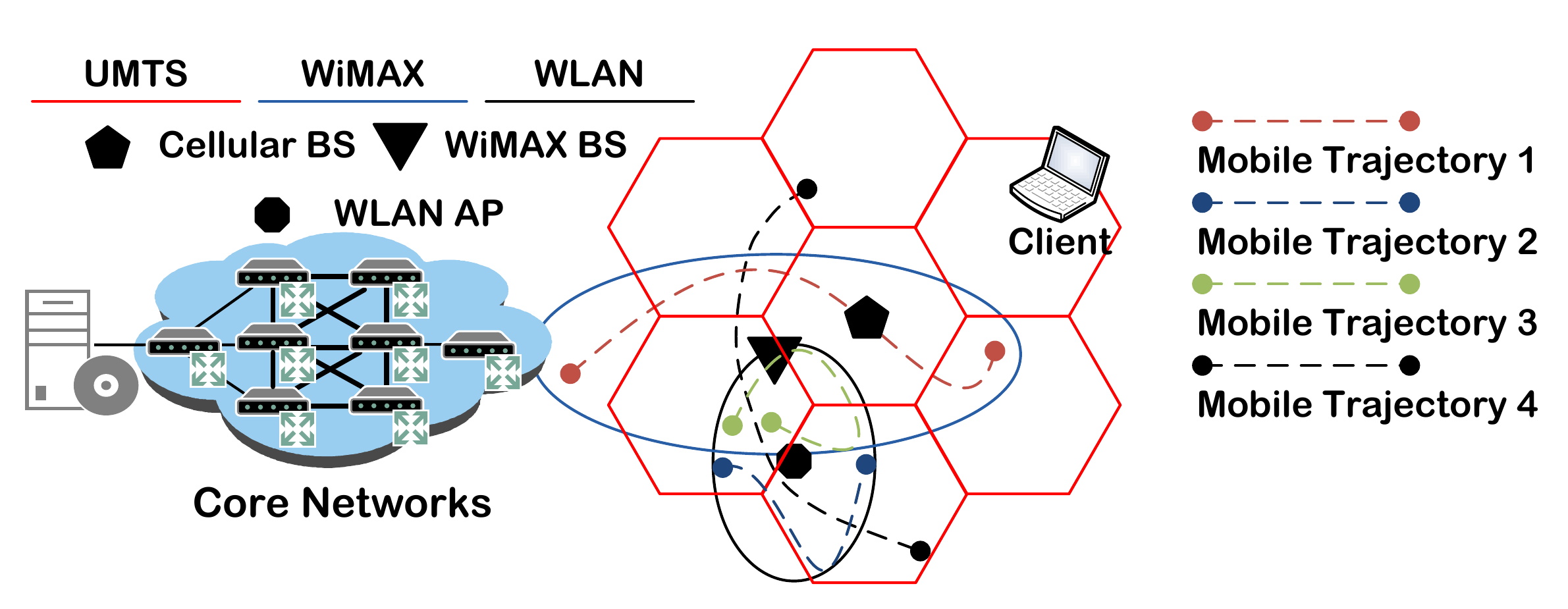}
 \caption{System architecture for performance evaluation.}\label{7}
 \vspace{-10pt}
\end{figure}
\begin{itemize}
    \item Exata 2.1 [32] is used as the network emulator. Exata is an advanced edition of QualNet [33] in which we can perform semi-physical emulations.
    \item H.264/SVC reference software JSVM 9.18 [35] is adopted as the video encoder. The generated video streaming is encoded at $30$ frames per second and a GoP consists of $8$ frames.
\end{itemize}

\subsubsection{Performance Metric}
We employ PSNR as the main metric of this study to evaluate the quality of the approach. PSNR (Peak Signal-to-Noise Ratio) is a standard metric of video quality and is a function of the mean square error between the original and the received video frames.

\subsubsection{Emulation Scenario}
We conduct all the emulations in the mobile scenario with trajectories indexed from $1$ to $4$ as shown in Fig. 4. The four mobile trajectories represent the different access options for the mobile user in the integrated heterogeneous wireless networks, e.g., the user could simultaneously access the UMTS and WiMAX while moving along the first trajectory. The mobile client requests to the server through a wireless interface and constructs the connection whenever it moves in the coverage. The moving speed of the client is set to be $2$ m/s in all the emulations.

For the confidence results, we repeat each set of emulations with different video sequences more than $5$ times and obtain the average results with a $95\%$ confidence interval. The microscopic and mobility results were presented with the measurements of finer granularity.
\subsection{Evaluation Results}

As shown in Fig. 8, PA-JSCC achieves higher PSNR values and lower variations than the other competing models. The average video PSNR in Trajectory $2$ is lower than that in Trajectory $1$ as the WLAN is less stable than the WiMAX. The results verify the instance in Fig. 1 and the conclusions in related work [6]. Besides, the superiority of PA-JSCC and FCVP over the other two schemes is larger in Trajectory $3$ and $4$ as more wireless access networks are available. The substantial improvements in video quality confirm the importance of JSCC in conjunction with flow rate allocation in heterogeneous wireless networks. PA-JSCC outperforms the FCVP as the Reed-Solomon code is more appropriate than the fountain code for the real-time video and thus reduce the erasure-coding-induced delays. In order to have a microscopic view of the results, we also depict the mean values and standard deviations (Stddev) of mobile trajectory $4$ in Table IV. The per frame video PSNR during the interval of $[0,\,20]$ second is presented in Fig. 9. It can be observed that PA-JSCC maintains the PSNR values at a relatively higher range. In the mobile trajectory $1$, the superiority of PA-JSCC over the JMRA becomes more obvious and due to the increase number of access options.
\begin{table*}[htbp]
\footnotesize
     \centering
         \renewcommand{\arraystretch}{1}
        \centering
      \caption{Average PSNR values for different videos.}
        \centering
        \begin{tabular}{|c||c||c||c||c||c||c||c||c|}
      \hline
        \multirow{2}{*}{Compared Approaches} & \multicolumn{2}{c||}{\emph{Foreman}} & \multicolumn{2}{c||}{\emph{Mother $\&$ Daughter}} & \multicolumn{2}{c||}{\emph{Hall}} & \multicolumn{2}{c|}{\emph{Container}}\\
       \cline{2-9}
        &\bfseries Mean & \bfseries Stddev  &\bfseries Mean & \bfseries Stddev  &\bfseries Mean & \bfseries Stddev&\bfseries Mean & \bfseries Stddev  \\
        \hline
        \hline
        PA-JSCC & $39.9$ & $0.62$ & $34.8$ & $0.51$& $35.7$ & $0.86$& $41.3$ & $0.93$\\
        \hline
      \end{tabular}
\end{table*}

\section{Conclusion and Discussion}
In this paper, we have presented a path adaption JSCC (PA-JSCC) approach for mobile video delivery in heterogeneous wireless networks. Through modeling and analysis, we have developed solutions for FEC redundancy adaption, video source rate adaption and flow rate allocation. Experimental results show that the proposed PA-JSCC is able to dynamically select the appropriate wireless access networks out of all candidates and significantly improve the video PSNR. As future work, we will consider: (1) design a seamless vertical handoff algorithm for optimal-quality video in the integrated WLAN, WiMAX and Cellular networks. The work in [5] formulates the heterogeneous wireless networks as a restless bandit systems. However, it does not provide in-depth analysis on the physical characteristics (e.g., the coverage and received signal strength) of each wireless network; (2) include an optimal path interleaving mechanism with the PA-JSCC for overcoming the burst loss.





\begin{thebibliography}{1}
\bibitem{1}
Hollywood, \emph{http://www.hollywood.com/.}
\bibitem{2}
Metacafe, \emph{http://www.metacafe.com/.}
\bibitem{3}
Cisco, ``Cisco visual networking index: Forecast and Methodology, 2012-2017,''\text{ } May 2013.
\bibitem{4}
T. Oliveira, S. Mahadevan and D. P. Agrawal, ``Handling Network Uncertainty in Heterogeneous Wireless Networks,'' in \emph{Proc. of IEEE INFOCOM}, 2011.

\bibitem{5}
P. Si, H. Ji, and F. R. Yu, ``Optimal network selection in heterogeneous wireless multimedia networks,''\emph{ Wireless Networks}, vol. 16, no. 5, pp. 1277-1288, 2009.
\bibitem{5}
J. Wu, B. Cheng, C. Yuen, Y. Shang, J. Chen, ``Distortion-Aware Concurrent Multipath Transfer for Mobile Video Streaming in Heterogeneous Wireless Networks,'' \emph{IEEE Transactions on Mobile Computing}, vol. 14, no. 4, pp. 688-701, 2015.
\bibitem{6}
S. Han, H. Joo, D. Lee and H. Song, ``An End-to-End Virtual Path Construction System for Stable Live Video Streaming over Heterogeneous Wireless Networks,'' \emph{ IEEE J. Select. Areas Commun.}, vol. 29, no. 5, pp. 1032-1041, 2011.

\bibitem{7}
P. Frossard and O. Verscheure, ``Joint source/FEC rate selection for quality-optimal MPEG-2 video delivery,'' \emph{IEEE Trans. Image Process.}, vol. 10, no. 12, pp. 1815-1825, 2001.

\bibitem{3}
J. Wu, C. Yuen, N.-M. Cheung, J. Chen, ``Delay-Constrained High Definition Video Transmission in Heterogeneous Wireless Networks with Multi-homed Terminals,'' to appear in \emph{IEEE Transactions on Mobile Computing}, DOI: 10.1109/TMC.2015.2426710.

\bibitem{8}
S. Ahmad, R. Hamzaoui, and M. Al-Akaidi, ``Adaptive Unicast Video StreamingWith Rateless Codes and Feedback,'' \emph{IEEE Trans. Circuits Syst. Video Technol.}, vol. 20, no. 2, pp. 275-285, 2010.

\bibitem{9}
M. Bystrom and J. W. Modestino, ``Combined Source-Channel Coding Schemes for Video Transmission over an Additive White Gaussian Noise Channel,'' \emph{IEEE J. Select. Areas Commun.}, vol. 18, no. 6, pp. 880-890, 2000.

\bibitem{10}
J. Wu, C. Yuen, B. Cheng, Y. Shang, J. Chen, ``Goodput-Aware Load Distribution for Real-time Traffic over Multipath Networks,'' to appear in \emph{IEEE Transactions on Parallel and Distributed Systems}, DOI: 10.1109/TPDS.2014.2347031.

\bibitem{10}
L. Qian, D. L. Jones, K. Ramchandran, and S. Appadwedula, ``A General Joint Source-Channel Matching Method for Wireless Video Transmission,'' in \emph{Proc. of the IEEE DCC}, 1999.

\bibitem{11}
X. Jaspar, C. Guillemot, and L. Vandendorpe, ``Joint Source-Channel Turbo Techniques for Discrete-Valued Sources: From Theory to Practice,'' \emph{Proc. IEEE}, vol. 95, no. 6, pp. 1345-1361, 2007.
\bibitem{11}
J. Wu, C. Yuen, M. Wang, J. Chen, ``Content-Aware Concurrent Multipath Transfer for High-Definition Video Streaming over Heterogeneous Wireless Networks,'' to appear in \emph{IEEE Transactions on Parallel and Distributed Systems}, DOI: 10.1109/TPDS.2015.2416736.

\bibitem{12}
Q. Xu, V. Stankovic, and Z. Xiong, ``Distributed Joint Source-Channel Coding of Video Using Raptor Codes,'' \emph{IEEE Trans. Circuits Syst. Video Technol.}, vol. 25, no. 4, pp. 851-861, 2007.

\bibitem{13}
Z. He, J. Cai, and C. W. Chen, ``Joint Source Channel Rate-Distortion Analysis for Adaptive Mode Selection and Rate Control in Wireless Video Coding,'' \emph{IEEE Trans. Circuits Syst. Video Technol.}, vol. 12, no. 6, pp. 511-523, 2002.
\bibitem{14}
J. Wu, B. Cheng, C. Yuen, N.-M. Cheung, J. Chen, ``Trading Delay for Distortion in One-Way Video Communication over the Internet,'' to appear in \emph{IEEE Transactions on Circuits and Systems for Video Technology}, DOI: 10.1109/TCSVT.2015.2412774.

\bibitem{14}
D. Jurca, P. Frossard, and A. Jovanovic. ``Forward Error Correction for Multipath Media Streaming,'' \emph{IEEE Trans. Circuits Syst. Video Technol.,} pp. 1315-1326, vol. 19, no. 9, 2009.

\bibitem{15}
J. Apostolopoulos and M. Trott, ``Path diversity for enhanced media streaming,'' \emph{IEEE Commun. Mag.}, vol. 42, pp. 80-87, 2004.
\bibitem{16}
J. Wu, C. Yuen, N.-M. Cheung, J. Chen, C. W. Chen, ``Enabling Adaptive High-Frame-Rate Video Streaming in Mobile Cloud Gaming Applications,'' to appear in \emph{IEEE Transactions on Circuits and Systems for Video Technology}.

\bibitem{16}
A. Ramaboli, O. Falowo, A. Chan, ``Bandwidth aggregation in heterogeneous wireless networks: A survey of current approaches and issues,'' \emph{Journal of Networks and Computer Applications}, vol. 35, no. 6, pp. 1674-1690, 2012.

\bibitem{17}
A. L. H. Chow, H. Yang, C. H. Xia, M. Kim, Z. Liu, H. Lei, ``EMS: Encoded Multipath Streaming for Real-time Live Streaming Applications,'' in \emph{Proc. of the IEEE ICNP}, 2010.

\bibitem{18}
V. Sharma, K. Kar, K. K. Ramakrishnan, and S. Kalyanaraman, ``A Transport Protocol to Exploit Multipath Diversity in Wireless Networks,'' \emph{IEEE/ACM Trans. Netw.}, vol. 20, no. 4, pp. 1024-1039, 2012.
\bibitem{19}
J. Wu, J. Yang, Y. Shang, B. Cheng, J. Chen, ``SPMLD: Sub-Packet based Multipath Load Distribution for Real-time Multimedia Traffic,'' \emph{Journal of Communications and Networks}, vol. 16, no. 5, pp. 548-558, 2014.

\bibitem{19}
K. Stuhlm\"{u}ller, N. F\"{a}rber, M. Link, and B. Girod, ``Analysis of video transmission over lossy channels,'' \emph{IEEE J. Select. Areas Commun.}, vol. 18, no. 6, pp. 1012-1032, 2000.
\bibitem{20}
J. Wu, Y. Shang, X. Qiao, B. Cheng, J. Chen, ``Robust Bandwidth Aggregation for Real-time Video Delivery in Integrated Heterogeneous Wireless Networks,'' \emph{Multimedia Tools and Applications}, vol. 74, no. 11, pp. 4117-4138, 2015.
\bibitem{20}
V. Paxson, G. Almes, J. Mahdavi, and M. Mathis, ``Framework for IP performance metrics,'' IETF, Tech. Rep. RFC 2330, May 1998.
\bibitem{21}
Y. Shang, J. Huang, X. Zhang, B. Cheng, J. Chen, ``Joint Source-Channel Coding and Optimization for Mobile Video Streaming in Heterogeneous Wireless Networks,''\emph{ EURASIP Journal on Wireless Communications and Networking}, pp. 1-16, 2013.

\bibitem{21}
N. Thomos, S. Argyropoulos, N. Boulgouris, and M. Strintzis, ``Robust transmission of h.264/avc video using adaptive slice grouping and unequal error protection,'' in \emph{Proc. of the IEEE ICME}, 2006.
\bibitem{22}
J. Wu, X. Qiao, Y. Xia, C. Yuen, and J. Chen, ``A Low Latency Scheduling Approach for High Definition Video Streaming in a Heterogeneous Wireless Network with Multihomed Clients,'' Multimedia Systems, vol. 21, no. 4, pp. 411-425, 2015.
\bibitem{22}
E. Baccaglini, T. Tillo, and G. Olmo, ``Slice sorting for unequal loss protection of video streams,'' \emph{IEEE Signal Process. Lett.}, vol. 15, pp. 581-584, 2008.
\bibitem{23}
J. Xiao, B. Cheng, Y. Shang, J. Huang, J. Chen, ``A Novel Scheduling Approach to Concurrent Multipath Transmission of High Definition Video in Overlay Networks,'' Journal of Network and Computer Applications, vol. 44, pp. 17-29, 2014.
\bibitem{23}
J. Xiao, T. Tillo, C. Lin, and Y. Zhao, ``Dynamic Sub-GOP Forward Error Correction Code for Real-Time Video Applications,'' \emph{IEEE Trans. Multimedia,} pp. 1298-1308, vol. 14, no. 4, 2012.
\bibitem{24}
Y. Shang, C. Yuen, B. Cheng, and J. Chen, ``TRADER: A Reliable Transmission Scheme to Video Conferencing Applications over the Internet,'' \emph{Journal of Network and Computer Applications}, vol. 44, pp. 161-171, 2014.
\bibitem{24}
M. Sun and A. Reibman, \emph{Compressed Video over Networks}. Marcel Dekker, Inc. New York, NY, USA, 2000.

\bibitem{25}
Exata, \emph{http://www.scalable-networks.com/exata}

\bibitem{26}
QualNet, \emph{http://www.scalable-networks.com/qualnet}

\bibitem{27}
JSVM, \emph{http://ip.hhi.de/imagecom-G1/savce/downloads/SVC-Reference-Software.htm}

\end{thebibliography}
%

\end{document}